\documentclass[12pt,halfline,a4paper]{ouparticle}
\usepackage[utf8x]{inputenc}
\begin{document}

\title{A survey on modelling of infectious disease spread and control on social contact networks}

\author{%
\name{Md Shahzamal}
\address{Macquarie University, Sydney, Australia}
\email{md.shahzamal@mq.edu.au}
\and
\name{Saeed Khan}
\address{University of Queensland,  Brisbane, Australia}
\email{s.khan@uq.edu.au}}

\abstract{Infectious diseases are a significant threat to human society which was over sighted before the incidence of COVID-19, although according to the report of the World Health Organisation (WHO) about 4.2 million people die annually due to infectious disease. Due to recent COVID-19 pandemic, more than 2 million people died during 2020 and 96.2 million people got affected by this devastating disease. Recent research shows that applying individual interactions and movements data could help managing the pandemic though modelling the spread of infectious diseases on social contact networks. Infectious disease spreading can be explained with the theories and methods of diffusion processes where a dynamic phenomena evolves on networked systems. In the modelling of diffusion process, it is assumed that contagious items spread out in the networked system through the inter-node interactions. This resembles spreading of infectious virus, e.g. spread of COVID-19, within a population through individual social interactions. The evolution behaviours of the diffusion process are strongly influenced by the characteristics of the underlying system and the mechanism of the diffusion process itself. Thus, spreading of infectious disease can be explained how people interact with each other and by the characteristics of the disease itself. This paper presenters the relevant theories and methodologies of diffusion process that can be used to model the spread of infectious diseases.}

\date{}

\keywords{diffusion; disease spread; contact networks, human movements, and covid-19}

\maketitle

\section{Introduction}
Infectious diseases are a significant threat to human society which was over sighted before the incidence of COVID-19, although according to the report of the World Health Organisation (WHO) about 4.2 million people die annually due to infectious disease. Due to COVID-19, more than 2 million people died during 2020 and 96.2 million people affected by this disease. The infectious diseases can also spread through animal contact networks, insect contact networks and even with plant contact networks~\cite{ keeling2008modeling, richardson2015beyond,jeger2007modelling}. Therefore, it is very important to understand the theories and methods for modelling spread of infectious disease and its control. The popular approach to model infectious disease spread is using theories and techniques of diffusion process on contact networks. The availability of high computational power allows researchers to simulate and analyse dynamics of infectious disease spread using the models of diffusion. Therefore, building simulation tools based on the diffusion modelling theories can help authorities to make policies and strategies to manage spreading of infectious diseases within a population.

A diffusion processes is defined as a dynamic phenomena on a networked system, connecting a set of nodes, that starts from a node or a set of nodes and spread over the networked system through inter-node interactions. In the diffusion process, contagious items (infectious particles, a piece of information, innovation and a specific behaviour etc.) initially grow on one or more nodes of the networks and then spread through neighbouring nodes over the network. The interactions among nodes are responsible for transmitting contagious items from one node to the other nodes. The interactions can happen through physical contact such as being in a common location, touch the same object within a time frame and transferring of objects from one node to other nodes through medium such as air. In infectious diseases  scenarios, infected individuals interact with susceptible individuals and disease is transmitted to susceptible individuals. Thus, the nodes of a networked system underlying of a diffusion process are represented by individuals and the inter-node interactions causing transmission of infectious particles are called contacts, edges, or links in the study of disease spreading on social contact networks~\cite{moss2016model,huang2016insights,shahzamal2017airborne}. 

Modelling of diffusion dynamics on networked system or social contact networks is required to understand what factors shape the diffusion behaviours. In reality, the factors governing the behaviours of diffusion process vary depending where it happens. However, the spreading of contagious items, e.g. infectious particles, over a contact network is often treated as the coupling of results of three factors, namely individual interactions, characteristics of contagious items, and environments. These three factors make strong contributions to diffusion phenomena on a contact network. The roles and influence of these factors depend on the context of diffusion processes.

Researches on diffusion modelling identified interactions between individuals are the main drivers for spreading contagious items on contact networks. The individual interaction patterns provide pathways to spread contagious items on contact networks. For example, how many susceptible individuals an infected individual meets during their infectious period determine the spreading speed of the infectious disease. If an infected individual is connected to many other individuals, there is a high probability to transmit disease to others by him. On the other hand, if an infected individual has no contact with other individuals, the disease is not transmitted. Thus, the final size of the epidemic, total number of infection caused during an outbreak of an infectious disease, depends on the contact patterns distribution in the population. There are several interaction properties such as how contact happens, contact frequency and contact duration etc. at the individual levels and contact degree distribution, clustering coefficient etc. at the network level that are studied to understand and model diffusion processes on social contact networks.

The contagious items can be infectious particles for spreading infectious diseases. The contagious items themselves play strong roles in their spread in a population and is identified as the critical factors to spread disease in the literature. The internal spreading potentiality of contagious items varies based on its characteristics. For example, the infectiousness of contagious particles is defined by the disease types and highly infectious diseases usually spread faster in a population. The spreading potential is also affected by the process of infectious particles generation. The impacts of contagious items also vary depending on the recipient individual behaviours. For example, the impacts of infectious particles varies according to the susceptibility of the individuals. Therefore, the characteristics of contagious items are often considered in the modelling of diffusion processes. In case of modelling infectious disease spreading, the characteristics of infectious items varies widely due to the various nature of disease.

The spreading of contagious items is often influenced by the environment. The environments represent the characteristics of the space where diffusion processes occur. For example, the impacts of infectious particles are determined by weather conditions such as temperature and humidity etc. The infectious particles generally lose their infectiousness over time and may also depend on the weather conditions. The underlying medium where contagious items spread can also be heterogeneous. For example, a disease can spread through multiple platforms such as proximity contact networks, transportation contact networks, and air-travel contact networks. Thus, diffusion modelling is also required to consider the heterogeneity of diffusion medium.

To model the diffusion on contact networks, these factors should be included to capture realistic diffusion behaviours. The impacts of contagious properties and environment have been studied for a long time in the literature~\cite{kim2018real}. However, analysing contact structure to understand the spreading of contagious items is comparatively new. The recent exploration of data on individual interactions have fuelled research on unravelling contact patterns affecting contagious spreading~\cite{shahzamal2019indirect, gonzalez2007complex,gonzalez2008understanding,barthelemy2005dynamical}. 

A spark has been seen on modelling of COVID-19 spread and control using social contact networks properties~\cite{bryant2020modelling, gatto2020spread, chang2020modelling, kuchler2020geographic}. The authors of ~\cite{bryant2020modelling} shows that there is significant variation in the number of connections and social contacts each individual make in a social interaction networks. Individuals with more social contacts are more likely to attract and spread infection. These individuals are likely the drivers of the epidemic, so-called superspreaders. When many superspreaders are immune, it becomes more difficult for the disease to spread, as the connectedness of the social network dramatically decreases.
Authors of ~\cite{thurner2020network} also show that properly taking some relevant network features into account, linear growth can be naturally explained. Further, the effect of nonpharmaceutical interventions (NPIs), like national lockdowns, can be modeled with a remarkable degree of precision without fitting or fine-tuning of parameters.

This paper analyse the approaches to integrate individual contact patterns with diffusion modelling and discuss how these approaches are used in modelling of infectious diseases.

\section{Diffusion modelling}
Diffusion modelling is an intensively researched area due to its wide applications. As the area of diffusion is diverse, the models developed are extremely varied in their approaches. Broadly speaking, the models developed can be divided into two groups based on their purposes: i) explanatory models, and ii) predictive models~\cite{meade2006modelling,guille2013information,pastor2015epidemic,barrat2008dynamical}. The explanatory models are usually developed to understand the factors affecting diffusion dynamics on a contact network. This often allows one to answer the questions such as which nodes are influential, what is the underlying reason for the way diffusion occurs, and what is appropriate diffusion controlling strategy? On the other hand, predictive models usually predict the spreading intensity and the final number of individuals received contagious items based on certain factors. It is often the case that the explanatory models find the key influential factors and apply them for developing predictive models. This paper focuses on understanding the diffusing approaches for explanatory models. There has been a range of approaches for explanatory models of diffusion. This section discusses some of them that are widely used in different fields of diffusion ranging from disease spreading on individual contact networks to information spreading on online social networks (OSN)~\cite{pastor2015epidemic,barrat2008dynamical}.

\subsection{Compartmental models}
Compartment epidemic models are frequently applied to study diffusion in many applications such as information diffusion, innovation diffusion and computer virus spreading~\cite{anderson1992infectious,himmelstein1984compartmental,gruhl2004information}. The fundamental concept of these approaches is to divide the system into several compartments or partitions. Each compartment represents a set of individuals having a specific status. Then, the dynamics of the diffusion are determined by the flows between these compartments. Widely used compartments are Susceptible having individuals who are exposed to the contagious items, Infected having individuals who have adopted the contagious items and started forwarding items (infecting) others, and Recovered having individuals who were infected but now recovered. This compartment model is called SIR propagation model and the dynamics of the compartments are given by
\begin{equation}
\frac{dS}{dt}=-\beta IS , \quad
\frac{dI}{dt}=\beta IS -\tau I , \quad
\frac{dR}{dt}=-\tau I
\end{equation}
where $S$, $I$ and $R$ are the fractions of the population in the Susceptible, Infected and Recovered compartments respectively, $\beta$ is the transmission rate from Susceptible to Infected compartment and $\tau$ is the transmission rate from Infected to Recovered compartment. Thus, the dynamics of the system is given by $S(t)+I(t)+R(t)=1$. In the diffusion modelling, $\beta$ is given as $\beta=\gamma \beta_{0}$, where $\gamma$ is the number of potential contacts on average individuals has with others through which contagious items transmit with a rate $\beta_{0}$. By changing the value of $\gamma$, $\tau$ and $\beta_{0}$ one, therefore, can study the diffusion dynamics of diffusion processes. 

The compartment model can be analysed easily mathematically in the simple case. It requires no more details than needed to reproduce and explain observed behaviours. It reduces data collection cost and computational cost. Clearly, it can be applied in many situations where high precision is not necessary. However, the assumption of having homogeneous interaction $\gamma$ between individuals is not realistic. Many researchers have pointed out that the interaction between individuals is clearly heterogeneous as individuals do not have the same level of contact with all its neighbours~\cite{mikolajczyk2008social,mossong2008social,hens2009mining,mossong2008social}. Moreover, the constant rate of transmission probability $\beta_{0}$ and constant recovery rate $\tau$ are not realistic in many diffusion processes. This is because individuals have different contact intensity with infected individuals and heterogeneous susceptibilities to the contagious items~\cite{wearing2005appropriate,smieszek2009mechanistic,smieszek2009models}. 

Heterogeneity is often integrated into the compartment model by dividing the main compartments (such as S, I, R) into sub-compartments. These sub-divisions can be constructed based on the age, risk behaviours, or spatial diversities of individuals. Then, the transmission probability can be divided into $k$ sub-classes and the model can be parameterised by means of a $k\times k$ transmission rate matrix instead of a constant transmission rate $\beta$~\cite{rhomberg2017parallelization, yorke1978dynamics}. For example, some disease spread models split individuals spatially (divide population with different regions) and assign heterogeneity for infection risks~\cite{lloyd1996spatial,pitcher2017network,meakin2018metapopulation}. These approaches are called meta-population models. Similarly, different infectious periods can be implemented in the model by dividing population into $k$ sub-classes which resolve the limitations of the constant rate of recovery~\cite{wearing2005appropriate}.

The integration of heterogeneity with sub-compartmentalisation relax some of the most unrealistic assumptions of basic compartmental models. However, many limitations of the compartment model still prevail and new issues arise by doing sub-compartmentalisation. The analysis in ~\cite{koopman2005mass} shows that individual's interaction is still random and transient in these models. Hence, individuals in the divided sub-populations behave homogeneously. In case of meta-population models, dividing a population into the various spatial groups can create asynchronous between these groups, but for time $t \rightarrow \infty$ they become homogeneous~\cite{keeling2005extensions}. Therefore, the real steady state heterogeneity cannot be captured by meta-population models. Thus, high partitioning in compartmental models lose the simplicity.  It also requires more data to fit the model and increases the data collection cost.

\subsection{Network models}
To overcome the limitations of the compartment models and capture realistic contact patterns, network science is adopted for modelling diffusion processes on the networked systems~\cite{newman2002spread,keeling2005networks}. The network based diffusion modelling is also empowered by the graph theory where contact networks are often generated by graph models. In addition, graph theory is applied to study the characteristics of contact networks. The core entities in the network-based modelling are nodes, representing individuals, and links connecting one node to other nodes in the network, which represent interactions between individuals. A contact network can be represented by a graph where vertices correspond to nodes and edges to links. The network models provide a range of flexibility for assigning nodes various attributes and defining links with a range of properties. A wide range of efficient network based models has been developed in the literature for studying diffusion processes~\cite{newman2002spread,keeling2005networks,holme2015modern}. For generating contact networks, a fundamental aspect is to build network structure called the network topology based upon which nodes interact with each other. There are four types of network structure namely regular lattice networks, random networks, small-world networks and scale-free networks that are frequently used to study diffusion processes~\cite{seyed2006scale, wang2003complex,hegselmann1998modeling,hegselmann1998understanding}. In addition, data-driven network models are also derived from real-world data and they may assume some properties of the theoretical models.

Regular lattices are the simplest representation of contact network structures where nodes are only connected to their nearest neighbour nodes in a lattice with a regular fashion ~\cite{hegselmann1998modeling,hegselmann1998understanding}. The regular lattice networks assume large path lengths, i.e. the average distance between two nodes is very high and the clustering coefficient is very high as well. Therefore, these are not realistic~\cite{sen2003small,zhou2005maximal}. The random network models improve regular lattice model where nodes contact with each other in a random fashion and each pair of nodes has an equal probability to be connected. Furthermore, the average path lengths in the random networks match with many real-world networks with appropriate contact probabilities~\cite{barbour1990epidemics}. However, the clustering coefficient is too low for these networks. Recently an approach is introduced called the small-world network model based on six degrees of separation phenomena which states that if you choose any two individuals anywhere on Earth, you will find a path of only six acquaintances on average between them. In a small-world network, most of the nodes are not neighbours of each other, but the neighbours of a given node are likely to be neighbours of each other. The Watts-Strogatz model~\cite{watts1998collective} generates such networks where the existing links of a regular lattice are re-wired with a defined probability. The generated networks assume high local clustering co-efficient and short path length~\cite{newman1999renormalization}. The latest approach to generate contact structure is scale-free networks developed by the Barabasi-Albert model~\cite{albert2002statistical} where node's degree follows a power-law distribution. In the Barabasi-Albert model, the scale-free networks are self-organised with growing and preferential attachment processes. The research found that many real-world systems have a power-law degree distribution~\cite{adamic2001search,clauset2009power}. 

The above structured contact networks can be analysed mathematically and numerically. The authors of ~\cite{morris2004network} has presented surveys on the methods of designing contact networks applying real contact data. These networks are, however, static in nature where node attributes and link properties are not changed during the observation period. These contact networks are often represented with adjacency matrices of binary values. This allows the use of algebra to calculate various network properties and corelate it with diffusion dynamics unfolded on it. While having some strong benefits over compartmental models, these network models still have several shortcomings, namely that in these models, the quality of contacts is overlooked. For example, the duration of contacts affects the transmission probability, and frequency of contact etc. There have, however, been some models to overcome these limitations with weighted contacts~\cite{onnela2007analysis,kamp2013epidemic,chu2009epidemic}. However, the weighted contact networks do not capture burstiness of the contact which is found to have an impact on spreading dynamics~\cite{lambiotte2013burstiness}. The other crucial temporal factor is that the contact sequences among individual are completely missing in these network models.

For studying diffusion processes with realistic contacts, there have been several approaches to make the contact network dynamic as well~\cite{holme2015modern,perra2012activity,shahzamal2018graph}. The dynamic networks assume the links are transient in status, i.e., links appear and disappear. However, the relationship between two linked nodes is often permanent. In the dynamic contact network models, the above static network models can be implemented as the underlying structure (capturing permanent social relationship among individuals) and an additional mechanism is added on top of that to maintain the link dynamic. The dynamic contact network models are often difficult to analyse with exact mathematical solutions. Thus, an approximation is often used to characterise the system dynamics~\cite{valdano2015analytical}. There are no analytical solutions for many dynamic contact network models and such models are only used for simulations to explore diffusion dynamics for wide scenarios of developed models. These classes of contact networks are often efficient tools to validate the simulations results of data-driven individual-level diffusion models.

\subsection{Individual-based models}
The other trend of diffusion modelling on contact networks is to apply individual-based models~\cite{bansal2007individual,cauchemez2008estimating,stehle2011simulation,toth2015role}. In these models, all operations are executed at the individual-level and thus the integration of many realistic contact properties becomes easier. The other fundamental concept is that individual-based models are implemented upon a community of targeted individuals and that are situated in an environment. In these models, every individual plays its role and interacts with its respective environment. Thus, infectious items are received by an individual according to its behaviour and surrounding conditions, and it transmits contagious items to other individuals by regenerating it. However, it is not so easy to define the boundaries of the model class based on individual compared to compartmental models or network models as the assumptions in the individual based models varies largely. 

The system dynamics in the individual-based model are generated with all individual actions happened simultaneously within the respective simulation environment of the individual~\cite{holland2006studying}. The respective environment depends on the modelling approaches and it may include parts or sometimes all of the other individuals. Thus, all individuals are affected by the state of neighbours in their simulation environment at the same time. Individual response to a specific environment can be deterministic or can be stochastic events. The reactions process to a simulation environment is often implemented with a set of rules (e.g. IF-THEN operations). For disease simulation, such a rule can be IF the individual is susceptible and if there was a contact with an infected individual, THEN switch the status from susceptible to infectious with a certain probability $P$. Some individual-based models implement the process of adaptation, learning or evolution. These models are called agent-based models, which is a subset of individual-based models, and have simulated intelligence~\cite{kiesling2012agent,copren2005individual, xu2017synthetic, stummer2015innovation}. 

The most significant advantage of individual-based models is that they allow for the inclusion of natural mechanisms for every desired aspect of the model to be as realistic as possible. They can offer characterisation even at the link level and environmental conditioning including complex biological mechanisms~\cite{keeling2008modeling,machens2013infectious}. The current exploration of data on social interactions leverages the benefit of this class of model as they easily allow modelling of individual interactions over time. The complexity of implementing higher-level architecture such as clustering and community structure are reduced as the network formation mechanism is implemented at a lower level. The contact networks created by dynamic contact network models can also be simulated with individual-level models. However, the individual based models become difficult with detailed information and require more effort to analyse sensitivity. To achieve stable insights, repeated simulations are conducted with a high number of parameter combinations. Therefore, the individual based model requires more computing resources and computation time. These models often cannot be analysed mathematically due to their stochastic nature and large number of parameters.

\section{Contact graphs for diffusion study}
Contact networks are widely used tools to study diffusion on social contact networks. This allows to characterise the diffusion behaviours and simulate the spreading dynamics. These contact networks can mimic the social interaction and reveal the influence of individual interactions. Contact networks are often generated by graph models. It is discussed in the previous section, dynamic contact networks are realistic to model diffusion processes. However, the development of dynamic graph models for generating dynamic contact networks is still at an early stage compared to the static models. There have been limited number of approaches to develop dynamic graph model. This section presents a brief details of current dynamic graph modelling approaches which consist of representing contact networks and links generation. Here, dynamic contact graphs/networks represent temporal or time-varying graphs/networks where edges between a pair of nodes are dynamic as their availability for transmission are not permanent.

\subsection{Dynamic graphs representation}
The evolution of a dynamic contact graph can be captured in many ways. The evolution in the graph can occur due to changes in the status of nodes and status of links. The links in the static graphs represent a relationship between a pair of nodes and is created there is at least one interaction during observation period~\cite{zhang2016modelling,starnini2013modeling}. In dynamic graphs, however, the links are often differentiated from contacts (links and contacts have different meaning)~\cite{holme2015modern,laurent2015calls,shahzamal2018graph}. The contacts indicate interactions between a linked pair of nodes occurring at certain times during an observation period. Dynamic graph modelling is required to incorporate timing information of these contacts with link dynamics. The development of a dynamic graph model often depends on how the graph is represented. The dynamic contact graphs can be represented in the following ways.

\textbf{Contact sequences:}
Many real world interaction data sets comes with the entries containing identities of interacted nodes and the time when the interaction happened, even with other some meta information such as gender and locations. The interaction time can be a time stamp  or a time interval sequence. For examples, works of ~\cite{mastrandrea2015contact,stehle2011simulation} have collected interactions between two individuals using RFID and wearable sensors. This representation is a straightforward and practical format computationally. However, analysing diffusion processes on the graphs with this format would be difficult as they do not count some properties such as contact duration. It is also difficult to visualise the contact graphs and hence representing it to audience.

\textbf{Multi-layer graphs:}
The dynamic graph can be visualised well if it is represented with as a sequence of static graphs. In this method, the observation time is divided into discrete time steps and a static graph is constructed for each time step~\cite{lee2015towards,kivela2014multilayer}. Thus, the dynamic graph becomes a multi-layer graph with each sequence of the static graph as a single layer. This allows one to understand and analyse the dynamic graph using static graph theories and then combine the results for the sequence of times to obtain overall results. This method is applicable where the time resolution is high (or continuous) compared to dynamic process on the studied graph. For studying infectious disease, this method has a limitation as the disease cannot be transmitted over a path during a sequence of the graph and thus analysing multi-layer graphs cannot capture real dynamics. The time-lines representation of contacts is one of the extended approaches where nodes are placed in one axis and times in another axis. The advantage with this representation is that the time-respecting paths (sequences of contacts of increasing times) between nodes are easy to identify as these are all paths that do not turn backwards in the time dimension. The structure of time-respecting paths can be represented as a binary matrix as it is in a static graph with an adjacency matrix. Thus, the dynamic graph can be expressed as a binary tensor. The limitation is that the corresponding adjacency tensor, as a data structure, takes a lot of memory and requires high computational overhead to process such graph ~\cite{valdano2015analytical,bach2013visualizing}.

\textbf{Dynamic links graph:}
In this representation, temporal variations are captured with only one dynamic graph where nodes and links change their status over time. The underlying graph is a static graph with the fixed links among nodes. In fact, the underlying graph captures the fixed topology of the dynamic graph and can be treated as the foot-print of nodes~\cite{holme2013epidemiologically,sarzynska2015null}. Then, the static graph structure evolves over time where contacts can appear and disappear. This means a time dimension is added with the static network. This approach is considered for the class of graphs where the targeted research question is to understand how the structure has evolved and how it affects the diffusion process unfolded on the graph. The dynamic graphs are typically data-oriented where the focus of study is on a data set, its structure, and how something behaves on it e.g. how disease spreading would behave on the graph. In addition, these observations may vary with used data sets and generalisation of results are often difficult. However, it is used widely due to its flexibility to implement and capture properties of real contact networks. 

\textbf{Time-node graphs:}
The recent trend of dynamic graph modelling is to extend the concept of node into temporal node i.e at each time step the same node is considered as a different node. Then, the graph is built among the temporal nodes. This approach is called the static expansion of a temporal graph~\cite{michail2016introduction}. This type of graph can be practical since it is straightforward to apply static graph methods also over the time dimension. Eventually one usually needs to map the time nodes back to the original nodes. This requires high computational power which is available in the current technology. However, the applicability is limited by the size of the networks.

\subsection{Dynamic graph modelling}
A general representation of the dynamic contact graph can be described as follows. Consider a dynamic contact graph $G_T$ that is built with a set of nodes $Z$, a set of relationships $L$ between these nodes (links, contacts), and a labelling sets $Y$ which represents any property such as links weights, set of node attributes; that is, $L \subseteq Z \times Z \times Y$. The relations between nodes are assumed to take place over a time span $\Gamma \subseteq \mathbb{T}$ denoting the lifetime of the system. The temporal domain $\mathbb{T}$ is generally assumed to be $N+$ for discrete-time systems or $R+$ for continuous-time systems. The dynamics of the system can be
subsequently described by a dynamic contact graph, $G_T = (Z, L, \Gamma, \phi, \psi )$, where\\
\[ \phi  : L \times \Gamma \rightarrow \{0,1\} \] $\phi$ is called presence function, indicates whether a given link is available at a given
time. The status of node can also be varied over time where they can be active or inactive to create links at a certain time. Thus, the model can be extended by adding a node status function \[ \psi : Z \times \Gamma \rightarrow \{0,1\} \]
where the activation function $\psi$ of nodes depend on a given time. Given a  $G_T = (Z, L, \Gamma, \phi, \psi )$, the graph $G = (Z, L)$ is called underlying graph of $G_T$. This static graph $G$ should be seen as a sort of footprint of $G_T$, which flattens the time dimension and indicates only the pairs
of nodes that have relations at some time in $\Gamma$. The connectivity of $G = (Z, L)$ does not imply that $G_T$ is connected at a given
time instant with the connectivity of $G$. Within this general graph definition, various contact networks can be generated by varying details in links and nodes definition. Thus, graph models are often accompanied with a network generation method. This network generation method handles incorporating details to links and nodes while graph defines the relationships at the abstract level. Some network generation methods that have been designed under the frameworks of various graph models are presented here.

\textbf{Static graphs with link dynamics:}
The simplest method of generating a dynamic graph is first to generate a static graph with links using a static graph model and then define a sequence of contacts for each link generated. To avoid complexity, contact generation process is often kept independent of the network position
of the links in this approach. The authors of ~\cite{holme2013epidemiologically} have applied the following procedures to generate a dynamic graph.

\begin{quote}
i) a static graph is constructed from a multigraph that is generated using the configuration model~\cite{albano2013matter} and deleting the duplicate links and self-links

ii) an active interval is generated for each link when contacts can occur i.e. nodes are present in the graph. The duration of active intervals is generated using a truncated power-law. For a link, the active interval starts at a uniformly random starting time within a sampling time frame (observation period)

iii) a sequence of contact times is generated following an inter-event time distribution over the observation period. This also generates burstiness in the contact patterns. 

iv) finally the contact time sequence is wrapped with the corresponding active interval of each link. The contact sequence times (step 2) which are within the active time interval are taken and other are deleted for a link. The wrapping is done for all links in the graph and a dynamic contact graph is obtained 
\end{quote}
The authors of ~\cite{sarzynska2015null} used a similar method to generate contact graph. These methods easily integrate contact dynamics and burstiness to the topology of a static graph. However, the inter-event times are not influenced by the topology structure and node properties. 

\textbf{Temporal exponential random graphs:}
The exponential random graph model (ERGM)~\cite{wang2009exponential} is widely used to generate static graphs in the study of social dynamics. The ERGM model parameters reflect the importance and weight of selected topological elements and sub-graphs such as triangles and stars. The ERGM model generates a class of graphs and the model parameter inferred from an empirical data capture the corresponding graphs. A similar modelling framework is used by the authors~\cite{hanneke2010discrete, krivitsky2012modeling} to generate temporal exponential random graphs (TERGM). Thus, the temporal dynamics of links are influenced by the network topology in TERGM. In TERGM, a set of states of the nodes observed over a time window for an ongoing dynamical process is applied to estimate the model parameters. Thus, the probability of making a contact between a pair of nodes is bounded with the time window of applied data set for modelling. The temporal exponential random graph models are not node-oriented model (nodes connectivity is not build at the node level, but connectivity is defined based on the network topology) that makes possible to change the network with its basic building block such as links. Thus, it is difficult to achieve a good fit unless the successive networks are close to each other~\cite{snijders2011statistical}.

\textbf{Coordinated temporal graph models:}
The above models cannot implement node and link level operations such as which neighbour nodes a host node contacts frequently, how recent changes in contact sequences affect the future contacts, whether the contact creation and deletion follow any specific social mechanism. There have been several works to incorporate these characteristics of dynamic contact graphs~\cite{cho2013latent, masuda2013self, stadtfeld2017dynamic, starnini2013modeling}. The work of ~\cite{starnini2013modeling} has developed a dynamic graph model for face-to-face interactions. This is a spatio-temporal graph implemented with a two-dimensional random walk. In this model, the propensity of walking closer to a node is proportional to the attractiveness assigned to it. Therefore, the
more attracted a walker is to its neighbour nodes, the slower its
walk becomes. A similar approach is implemented in the dynamic graph model developed by ~\cite{mantzaris2012model}. This model for online setting and their assumption is that some individuals are much more central in a temporal graph than they are in an aggregated static graph. Thus, random communication partners are assigned to a node by a basal rate and a positive feedback mechanism. The authors have applied stochastic point processes to model dynamic contact graphs. In this model, a node creates and breaks links according to a Bernoulli process with memory. The probability of an event between two nodes increases with the number of events recently occured between them.

\textbf{Activity-driven graph modelling:}
A comparatively simple dynamic graph modelling approaches is proposed by~\cite{perra2012activity}. This model is called activity driven network modelling (ADN). They adapt the graph sequence
framework of dynamic graph modelling and generate a simple graph $G_T$ at
(it is a discrete time system) time $t$. The graph generation procedures are as follows:

\begin{quote}
 i) node $i$ is assigned an activity potential $a_i$. This is usually done with a power-law distribution. The activity potential is assigned to all nodes in the graph

ii) the graph generation process goes through increasing a time counter to $t$ and assume that $G_T$ is empty, i.e. all $N$ nodes have no link and contact memory from previous time step

iii) activate node $i$ with a probability $a_{i} \Delta t$. If node $i$ is activated, it is connected with other $m$ randomly chosen distinct nodes. Repeat the step 2 and step 3
\end{quote}

The distinguishing characteristics of this model are that the activity of the nodes governs the link creation. In contrast, the previous models are connectivity driven where the network's topology is at the core of the model formation. The ADN overcomes the timescale separation assumption and explicitly accounting for the concurrent evolution of the interactions in a graph and the dynamic process evolving on it~\cite{rizzo2016innovation}. The studies of ~\cite{gomez2011nonperturbative,sun2015contrasting} show that many important aspects of the system dynamics can be characterised using a heterogeneous mean-field approach. Interestingly, it is found that some system properties are directly related to the activity potential of nodes. The diffusion dynamics are quite different from that of aggregated static networks~\cite{sun2015contrasting}. However, the basic ADN model has several limitations such as nodes contact with a fixed number of links during each activation, the contacts are not repetitive, and social structure among individual is not maintained. 

\subsection{Realistic activity driven graph models}
When nodes are active in the basic activity driven network, they randomly create connections with other nodes. In other words, at any time, a node may connect with any other nodes in the graphs. That means this method does not apply any prior knowledge of social or geographic relationships that could alter the selection of one link over another. However, this assumption severely challenges the feasibility of the ADN modelling when it is implemented for real graphs. There have been great efforts to incorporate realistic features with the basic ADN. An individual can interact with other individuals arbitrarily or by choice. Thus, some links tend to be persistent in time and such links are generated in the household, at the office and with close friends. This property is integrated introducing memory effects in the link formation. Therefore, social graphs can have two types of links. The first class describes strong ties that identify time repeated and frequent interactions among specific couples of nodes. The second class characterises weak ties among agents that are activated only occasionally. It is natural to assume that strong ties are the first to appear in the system, while weak ties are incrementally added to the contact set of each node. This approach is studied in the work of ~\cite{karsai2014time} where it is assumed that a node will connect to a new node with a probability $P(n+1)=\frac{\eta}{\eta+n}$, where $n$ is the current contact set sizes of the node and $\eta$ is the tendency to broaden their contact set sizes. Therefore, the probability of contacting with a node from previously contacted nodes is $1-P(n)$. This method of repeating with the old contacted nodes and extending contact set size is called the reinforcement process. In the above process, every node has the same tendency to extend the contact graph. In reality, however, individual have heterogeneous tendency to extend the contact set size. This issue is addressed by the work of ~\cite{ubaldi2017burstiness} where they proposed to assign a heterogeneous value of $\eta$. For a node $i$, the probability of contacting a new node is given by
\begin{equation*}
    p(n_i)= \left(1+\frac{n_i}{\eta_i}\right)^{-\alpha _i}
\end{equation*}
where $\alpha_i$ is the reinforcement of node $i$ and $\eta_i$ is the characteristic number indicating the size of contact set size before reinforcement start. The value of $\eta_i$ is often assigned with power-law and the distribution of resultant contact set size will be a power-law. 

In the social graphs, individuals have a tendency to make a close social circle and make a community. Therefore, the underlying social structure of a dynamic contact graph also should maintain community structure. The community structure emerges in a graph by creating triadic closure when making a new connections~\cite{bianconi2014triadic}. One typical mechanism to make triadic closure is to use common neighbours (CN) indices, where two nodes $i$ and $j$ are going to interact if their neighbour nodes set has substantially overlap. This means that the probability of these two nodes interacting is proportional to the number of common neighbours. In other words, triadic closure can be created if the host node chose a new neighbour from its neighbour's neighbour~\cite{bianconi2014triadic,daminelli2015common}. The random new neighbour selection mechanism of basic activity driven graph can not emerge the community saturate. The work of ~\cite{laurent2015calls} has upgraded the basic ADN integrating a triadic closure creation mechanism. If a host node $i$ has not contacted any node yet, it randomly picks another node from the entire graph $j$ and creates a link. Otherwise, the host node tries to make a new link with the triadic closure mechanism. As the first step, it selects randomly one neighbour node $j$ from his contact set with a probability. If node $j$ is not selected or has no other neighbours node except node $i$, node $i$ looks for another random new node and creates a link. If node $j$ is elected and has neighbour nodes, then it selects a random neighbour node $k$ from the neighbour set of $j$. Then node $i$ from a link with $k$ and create triadic closure.

The basic activity-driven graph model assigns heterogeneous potentiality to nodes. This generates a heterogeneous distribution of interactions. However, the research on social interaction shows that individual's interactions have bursty nature, i.e the inter-event time of activation of nodes in ADN is required to be heterogeneous~\cite{stehle2010dynamical,lambiotte2013burstiness}. This bursty activity has a strong influence on graph evolution and diffusion unfolded on it. The inter-event time $t_i$ is directly connected with the activity of node $i$ and can defined as $a_{i}=\frac{1}{<t_i>}$. The inter-event time usually spans over several orders of magnitude. The authors of ~\cite{ubaldi2016burstiness} shows a mechanism to capture this bursty nature of human dynamics using a power-law distribution. The basic ADN can generate heterogeneous contact degree distribution based on the value of $a_i$. However, the growing of contact set size is completely dependent on $\eta$. In addition, the model can capture link heterogeneity of directed graphs where the in-link propensity can be different to out-link density. This issue is addressed by the work of ~\cite{pozzana2017epidemic} where each node is assigned an attractiveness. When a node selects a new neighbour node, a node will be chosen based on its attractiveness. However, any current modification of ADN can generate contact networks with links for indirect interactions. The study of ~\cite{shahzamal2018graph} shows how to add repetitive contacts and indirect contacts in generating contact networks.

\section{Infectious disease diffusion}
The theories of diffusion process can be applied to model infectious disease spreading. It is required to understand how the infectious disease spreading is modelled. This allows to integrate the above diffusion model with the disease spread. In the infectious disease, two key aspects are disease transmission and calculation of transmission force. 

\subsection{Disease transmission}
An infectious disease is transmitted from an infected individual to susceptible individuals via transferring organisms/microbes capable of causing infection~\cite{fernstrom2013aerobiology,boone2007significance}. These organisms/microbes are called pathogens. In this paper, contagious items or infectious particles refer to these pathogens. The infectious items enter the body of susceptible individuals and deposit on mucus membranes of body parts such as mouth, nose, throat, and lungs where they can cause an infection. Therefore, for an infectious disease to persist within a population, relevant contagious items are required to be transmitted continuously to new bodies. The contagious items are transmitted through two mechanisms: 1) direct transmission and 2) indirect transmission~\cite{shahzamal2019indirect, shahzamal2018impact, shahzamal2017airborne, shahzamal2018graph, shahzamal2020vaccination}. Direct transmission occurs through individual-to-individual interactions transferring contagious items without any intermediate transmission medium between these two individuals. The example includes physical touches (such as shaking hands, kissing etc.) and contact of blood and body fluids. Direct transmission is found in infectious disease such as common colds, sexually transmitted diseases etc. For many infectious diseases~\cite{rottier2003controlling,brankston2007transmission,fernstrom2013aerobiology,boone2007significance}, infected individuals generate particles containing infectious microbes by their respiratory activities like talking, laughing, coughing or sneezing. These particles are scattered into the environment of the proximity of the infected individuals. The infectious particles then deposit onto objects or surfaces and survive long enough time to transfer to other susceptible individuals who subsequently touch the objects. This creates the indirect transmission of diseases where intermediate medium or objects are required to transmit infectious particles. Examples of diseases with the indirect transmission are Coronavirus, Rhinovirus, and Influenza etc.

The ways indirect transmission occur are not the same in all cases and can be classified into different modes which are based on the roles of the intermediate medium and the properties of the infectious particles when transmitting through the intermediate medium. The respiratory activities of infected individuals generate droplets containing infectious particles and the sizes of the droplets often define the mode of transmission. Droplet whose size is comparatively large, often assumes to be greater than 5$\mu$m, are transmitted through the air to nearby susceptible individuals. This mode of indirect transmission is called droplet transmission. However, the droplet whose size is small, often assumes to be less than 5$\mu$m, evaporates quickly and becomes droplet nuclei. These droplet nuclei are suspended in the air for a long time and can travel large distances. Thus, they can transmit to susceptible individuals with a long time delay after their generation, even to susceptible individuals who are far away up to 100m from the source infected individual. This mode of indirect transmission is called airborne transmission. Indirect transmission of infectious particles can also happen through vectors (mosquitoes, flies and mites etc.) that carry infectious particle from an infected individual to susceptible individuals with delay and at substantial distances. In this situation, the contagious items present in the blood or skin of an infected individual are ingested by vectors. Then, it is developed in the vectors itself. Susceptible individuals are usually infected through the bite of an infectious vector, though other ways of entry are possible. Examples of vector-borne disease transmissions are yellow fever, malaria, plague and dengue etc.~\cite{rottier2003controlling,brankston2007transmission,fernstrom2013aerobiology,stoddard2009role}. Another indirect mode of transmission is to spread disease through contaminated objects spatially. Examples of such diseases includes water-borne diseases, food-borne diseases~\cite{brennan2008direct,lange2016relevance}.

It has been observed that the spreading of some infectious diseases is dominated by only direct transmission and can be modelled by creating direct transmission links for co-presence interaction between infected and susceptible individuals. However, the infectious diseases that spread based on indirect transmission or have additional indirect transmission along with direct transmission cannot be modelled by creating only direct transmission links for co-presence interactions. This thesis considers airborne infectious disease spreading as a case study for understanding and modelling the impacts of indirect transmission links. For airborne diseases, infected individuals generate droplets containing infectious particles through various respiratory activities. The authors of ~\cite{lindsley2012quantity} have found that an infected individual generates on average 75,000 particles/cough but it can be up to 500,000 particles/cough. They have also found that 60\% of these particles can reach the alveolar region of lungs if the particles are inhaled by another individual. It is found that the cough frequency of an infected individual is on average 18/hr~\cite{loudon1967cough}. Thus, an infected individual deposits about $1.36\times10^{6}$ particles during a one hour stay at a location. Up to 50\% of these particles evaporate and become droplet nuclei (airborne particles) which are suspended in the air for a longer time~\cite{thomas2013particle}. Airborne particles are also added to the environment by breathing, talking and laughing. There have been a wide range of studies to understand the viral load of airborne particles. The studies show that most of the influenza virus is contained in the droplets whose sizes are $<5\mu$m. The works of ~\cite{lindsley2010measurements,lindsley2015viable,lindsley2010measurements} have found that up to 75\% virus is contained within droplets with sizes $<5\mu$m. The exhaled breath of an influenza patient can generate on average 0.5 plaque-forming units (PFU) for influenza viruses~\cite{nikitin2014influenza}. The study of ~\cite{lindsley2015viable} found that a cough can generate up to 77 PFU virus. Inhalation of 0.7 - 3.5 PFU of influenza is sufficient to cause infection in 50\% of susceptible individuals~\cite{alford1966human}. Therefore, it can be concluded that the generated airborne particles have sufficient viral load to cause infection if they are inhaled.

The impact of airborne transmission is different to the other model of indirect transmission as airborne particles can travel spatially while large droplets settle nearby. The literature indicates the various range of travel distances for airborne particles. The travel distances depend on the weather conditions and air-flows. The authors of ~\cite{han2014risk} show that airborne particles can travel up to 100m in the direction of air-flow. The travel distance can also be interpreted from the analysis of SARS outbreak occurred in the Amoy Gardens Hong Kong in the year 2003. The study of~\cite{yu2004evidence} has revealed that the infection had reached the Block-E which was at 60m distance from the Block-B where infection had started, although there was no indication of physical interaction among the residences of these buildings. Thus, it was concluded that the infection particles travelled to the Block-E through airborne transmission. A number of studies have also shown that it is also possible to disperse airborne particles between flats in a building~\cite{mao2015airborne,gao2008airborne} and between wards in a hospital~\cite{beggs2003airborne}. The experiment of ~\cite{gao2008airborne} shows that airborne particles can also travel from one building to the nearby buildings. The travel distance of airborne particles is extended in the open area. The authors of ~\cite{corzo2013airborne} have studied the presence of influenza A virus around pigs farms by collecting air samples at different distances from the farms. They have noticed a significant amount of RNA copies of the virus at the distance of 1.5Km from the pig farms that had influenza A infected pigs. Therefore, the airborne indirect transmission mode of infectious disease has strong potential to spread diseases. The airborne infectious diseases spreading is an important application of diffusion process with indirect transmissions.

\subsection{Infection risk}
An interaction (e.g. being in the same location) between infected and susceptible individuals poses an infection risk for the susceptible individual. Infection risk assessment can be divided into two steps: determining the intake dose of infectious particles and finding the corresponding infection probability~\cite{sze2010review}. The infectious particles that reach the target infection site are called the intake dose. The intake dose is estimated based on the exposure level to the infectious particles, the pulmonary ventilation rate of susceptible individuals, the exposure time interval, and the respiratory deposition of the infectious particles. Then, the infection probability is calculated by a mathematical formula. Two approaches are applied to determine if an infection occurs: deterministic and stochastic. The first approach assumes that each individual has an inherent resistance up to a dose of infectious particles. Thus, a susceptible individual contracts the disease when a target infection site is exposed to a dose equivalent to or exceeding the threshold dose. In the stochastic approach, any amount of intake dose causes disease with a certain probability. The infectious particles are usually randomly distributed in the suspension medium. Thus, the estimated exposure level and intake dose of airborne particles are always expected values rather than exact values. Therefore, the stochastic models are appropriate for studying airborne disease spread. Models that are frequently used for assessing infection risk for airborne diseases are now discussed.

A wide range of models has been developed for the spread of airborne disease. These range from simple models that are easy to apply to complex models that require greater detail of the disease spreading process. Unfortunately, these details is not always available for many diseases. In the literature, the Wells-Riley model or its modification are widely used to estimate infection risks~\cite{sze2010review,fennelly1998relative}. The Wells-Riley equation is given as  
\begin{equation}
P_I=1-exp\left(-\frac{Igpt}{Q}\right)
\end{equation}
where $P_I$ is the probability of causing infection to a susceptible individual for the intake dose $E=Igpt/Q$, $I$ is the number of infected individuals at the interaction room, $p$ is the breathing rate of the susceptible individual (L/s), $g$ is the average quanta generation rate (quanta/s), $t$ is the exposure time interval, and $Q$ is the room ventilation rate (L/s). The $P_I$ is, in fact, the ratio between the number of infections caused for $E$ and the susceptible individuals. This model is based on the concept of quanta which is the number of droplet nuclei required to cause infection for $63\%$ of all exposed susceptible individuals. The ratio $P_I$ provides the reproduction number of the studied diseases which is frequently used to determine disease spreading dynamics for the large population. The model parameter quanta generation rate $\vartheta $ is required to be estimated from the real outbreak cases. This is very difficult for many diseases as it requires data from real outbreak scenarios. The model is also limited due to its assumption that particles are homogeneously distributed in the air, and that every particle reaches to the target infection site. It does not consider the duration of particle generation.

There have been several modifications to overcome these limitations. The authors of ~\cite{fennelly1998relative} incorporated the effect of respiratory protection system that may filter the inhaled infectious particles by multiplying a fraction term with the intake dose as $E=\frac{Igpt \theta}{Q}$, where $\theta$ is the fraction of infectious particles reached to a target infection site. Air disinfection and particle filtration are used in the many buildings that reduce the effective infectious particles to cause infection. These factors are included for the Wells-Riley equation in the work of ~\cite{nazaroff1998framework}. However, collecting such data is difficult and expensive for large scale simulation. The assumption about the homogeneity of particle distribution in the interaction area is addressed by ~\cite{gammaitoni1997using}. They considered the time-weighted average pathogen concentration in the room air to incorporate the non-steady-state conditions in the Wells-Reily equation. This model is given by
\begin{equation}
P_I=1-exp\left(-\frac{pIg (Qt+e^{-\varphi t}-1)}{VQ^2}\right)
\end{equation}
where $Q$ is the air change rate or disinfection rate, $\varphi $ is the particle accumulation rate and $V$ is the volume of interaction area. In spite of these improvements, the Wells-Reily model still requires the total exposure during an outbreak to find the quanta generation rate and that is not possible for many diseases. 

Rudnick and Milton~\cite{rudnick2003risk} developed a model where the exhaled air volume fraction is used to estimate the number of quanta that the susceptible individuals are exposed to:
\begin{equation}
P_I=1-exp\left(-\frac{ g \bar \omega It}{N}\right)
\end{equation}
where $\bar \omega $ is the average volume fraction of room air that is exhaled breath and $N$ is the total number of people in the premises. To find the quanta generation rate based on the $\bar \omega $, one requires a knowledge of carbon dioxide concentration in the room. These models still follow the well-mixed assumption of particle concentration. Some works~\cite{gao2008airborne,tung2008infection} address this problem by experimenting the dispersion of tracer gas and integrating impacts with model.

In the models discussed above, the quanta generation rates are not well understood for many diseases. However, the infectious particles generation rates, their formation, pathogen loads and their survivable time etc. are now becoming available. The authors of~\cite{nicas1996analytical} first introduce a dose response model based on the infectious particles concentration instead of quanta. The model is 
\begin{equation}
P_I=1-exp\left(-\frac{Ig\theta pt}{Q}\right)
\end{equation}
where $g$ is the number of infectious particles released per infected per unit time and $\theta$ is the fraction of infectious particles reaches the target site. In this equation, the quanta generation rate $g$ is replaced by $\theta g$. The authors defined the source strength $g$ with cough frequency, pathogen concentration in the respiratory fluids and the volume of expiatory droplets introduced into the air in a cough. This model also based on the homogeneity. Recently, the authors of ~\cite{issarow2015modelling} have also introduced a model based on the infectious particle concentration considering non-steady-state conditions as
\begin{equation}
P_I=1-exp\left(-\frac{I(g -r)\theta p t}{Q}\left[ 1-\frac{V}{QT}\left(1-e^{-\frac{Q \tau }{V}}\right)\right]\right)
\end{equation}
where g is the particles generation rate, $r$ is the mortality rate of the generated particles, $\theta$ is the deposition fraction of the inhaled particles, $\tau$ is the duration of particle generation, and $t$ is the duration susceptible individuals breath in infectious particles.

In the above equations, the temporal variation in the particle concentration is captured using a non-steady-state model. However, this model assumes that all infected individuals arrive at the same time and this may not happen in reality. The variations in the arrival time of infected individuals also introduce the fluctuations in the particles concentration. The current models also do not capture the exposure that susceptible individuals receive after infected individuals leave the interaction locations. Thus, it would be more appropriate to find exposure level due to contact with each infected individual and sum them up to find total exposure. Therefore, the arrival and departure of each infected individual can be tracked independently and hence the exposure during indirect interactions. This also allows one to assign a random value of $Q$ to each contact to capture heterogeneous particle concentrations at different locations.

The disease transmission probability can also be calculated for both the direct and indirect contacts as follows~\cite{shahzamal2019indirect}. This model resolves the problems mentioned in the previous section. This model define a contact called same place different time (SPDT) that has both direct and indirect infection transmission probability. This accounts when an infected individual and susceptible individuals have been to a place. If a node in the susceptible compartment receives a SPDT link from a node in the infectious compartment, the former is subject to exposure $E_l$ of infectious pathogens for both direct and indirect transmission links according to the following equation
\begin{equation}
E_l =\frac{gp}{Vr^2}\left[r\left(t_i-t_s^{\prime}\right)+ e^{rt_{l}}\left(e^{-rt_i}-e^{-rt_l^{\prime}} \right)\right]
+\frac{gp}{Vr^2}\left(e^{-rt_l^{\prime}}-e^{-rt_s^{\prime}} \right)e^{rt_{s}}
%\end{split}
\end{equation}
where g is the particle generation rate of infected individual, p is the pulmonary rate of susceptible individual, V is the volume of the interaction area, r is the particles removal rates from the interaction area, $t_s$ is the arrival time of the infected individual, $t_l$ is the leaving time of the infected individual, $t_s^{\prime}$ is the arrival time of susceptible individuals and $t_l^{\prime}$ is the leaving time of susceptible individuals from the interaction location and $t_i$ is given as follows: $t_i=t_l^{\prime}$ when the SPDT link has only a direct component, $t_i=t_l$ if the SPDT link has both direct and indirect components, and  $t_i=t_s^{\prime}$ otherwise. If $t_s<t_s^{\prime}$, $t_s$ is set to $t_s^{\prime}$ for calculating an appropriate exposure~\cite{shahzamal2019indirect}. If a susceptible individual receives $m$ SPDT links from infected individuals during an observation period, the total exposure $E$ is 
\begin{equation}\label{eq:expo}
E=\sum_{k=0}^{m}E_{l}^{k}
\end{equation}
where $E_{l}^{k}$ is the received exposure for k$^{th}$ link. The probability of infection for causing disease can be determined by the dose-response relationship defined as 
\begin{equation}\label{eq:prob}
P_I=1-e^{-\sigma E}
\end{equation}
where $\sigma$ is the infectiousness of the virus that causes infection~\cite{fernstrom2013aerobiology}.

\section{Diffusion control}
\subsection{Diffusion control strategies}
Controlling diffusion dynamics on individual contact networks has a wide range of applications ranging from  mitigating the spread of infectious diseases to marketing products. The methods developed for controlling diffusion depends on the context and applications. For example, controlling diffusion for marketing a products focuses on maximising the spreading of items to the largest proportion of populations~\cite{van2007new, peres2010innovation} while diffusion controlling for infectious disease focuses on minimising the number of infections reducing the number of individuals received spreading items~\cite{yang2016optimal}. However, the key task in all controlling methods is to find a set of individuals and change their behaviours to alter spreading rates. These individuals often have high spreading potential and are called super-spreaders. The size of the set should be minimal to reduce vaccination cost as well as achieve the control goals. 

Most of the efforts of developing a control strategy are put on finding the optimal set of individuals. Accordingly, researchers search for the contact properties of individuals and their behaviours relevant to the spreading of contagious items~\cite{al2018analysis,scholtes2016higher,kas2013incremental}. Individual's preference and exposure intensity to contagious items, personal status and their surrounding environment often define the spreading potential. Thus, the influential individuals are often searched based on the individual's behaviours~\cite{ma2016rumor,ma2018rumor}. The network properties such as the number of connection of individuals to others are also key factors to determine one's spreading potential. Understanding personal behaviours and modelling is a complex process. In addition, research focuses on understanding the impacts of contact properties on diffusion dynamics.

Various measures of network properties are applied to find influential individuals for controlling diffusion of contagious items. The widely used measures to find the important individuals are degree centrality, betweenness centrality, k-core score and PageRank centrality etc~\cite{al2018analysis}. Individuals contact degrees defined the number of connections to other individuals is frequently used as topological measures of influence. In the social contact networks with broad degree distribution, it is observed that individuals with high degree determine the diffusion dynamics~\cite{albert2000error}. However, degree based methods sometimes underestimate the low degree individuals that can be influential through connecting high degree individuals. The page ranking algorithm developed to rank the content in the World Wide Web is also adapted to find the pivotal individuals in social contact networks. The ranking mechanism of PageRank is simple and straight forward~\cite{brin1998anatomy} where the importance of page is measured by counting the number and quality of links to that page. The PageRank algorithm is only applicable to directed networks. Betweenness centrality is also a good candidate in many applications as it is a measure of the number of shortest paths passing through one individual ~\cite{freeman1978centrality}. Thus, it is most likely that individuals having high betweenness centrality will play key roles in shaping the diffusion dynamics on the networks. It is efficient but requires high computational resources. Moreover, it is applicable for undirected networks. The K-core score present the positions of individuals in the social networks with the k-index obtained by iteratively removing k-degree nodes~\cite{wuchty2005evolutionary}. These networks measures are based on static networks and may not provide optimal performance in dynamic contact networks. 

The one way of finding influential nodes in dynamic contact networks is to use temporal versions of traditional centrality metrics~\cite{scholtes2016higher,kas2013incremental}. In this work, time respecting paths, paths are created based on the time order of links availability, are used to calculate the betweenness centrality and closeness centrality. Eigenvector centrality is also modified for dynamic networks~\cite{taylor2017eigenvector}. These methods require complete information regarding contact networks. The random walk is applied in ~\cite{rocha2014random} for measuring temporal centrality. This does not require global information. However, all these algorithms require huge computational resources. The temporal centrality measurement is in the early stage and is not still feasible for applying in large social contact networks. 

Applicability of these methods depends on the application scenarios. For example, betweenness centrality can be applied to find the influential individuals in online social contact networks as the contact information often is available~\cite{al2018analysis,newman2005measure}. 

\subsection{Disease spread control}
It is difficult to apply the above approaches for controlling disease spreading as it is quite difficult to collect contact information of a population. The vaccination strategies are required to be developed based on the contact information that can be obtained locally. The key task of a vaccination strategy is to choose a set of individuals based on the local contact information. The simplest way of selecting a set of individuals is to choose randomly from the population and is called random vaccination~\cite{madar2004immunization,cohen2003efficient,lelarge2009efficient}. This approach does not consider the disease spreading behaviours of the chosen individuals. Therefore, the information collection cost is minimal. However, it requires a large set of individuals to be vaccinated for achieving hard immunity to disease. Thus, the research is directed to select the individuals who have strong disease spreading potentials~\cite{pastor2002immunization}. These methods are called targeted vaccination. In the targeted vaccination, the number of individuals to be vaccinated is often small and the effectiveness of strategies is substantially high if an appropriate set of individuals are chosen. Therefore, the infection cost can be substantially lower in a targeted vaccination strategy with the reasonable cost of information collection and vaccination cost.  

There has been a wide range of vaccination strategies using obtainable contact information~\cite{deijfen2011epidemics,britton2007graphs,mao2009efficient, kucharski2020early}. All these methods do not develop vaccination strategies based on local contact information. Sometimes global information is also used and methods are developed to find the global metrics with locally obtainable contact data. There have been several other methods that depend on the movement behaviours of individuals instead of collecting information on interactions between individuals. Vaccination strategies are also varied based on the implementation scenarios. There are two specific vaccination scenarios: preventive vaccination (pre-outbreak) and reactive vaccination (post-outbreak). The vaccination strategies that apply local contact information is now first discussed. Then, the implementation of vaccination strategies is discussed. As examining vaccination strategies in the real-world scenarios are expensive and difficult, empirical contact networks or synthetic contact networks are frequently applied to test and validate developed vaccination strategies. The discussion includes the network model based vaccination strategies.

The authors of~\cite{cohen2003efficient, britton2007graphs} present an elegant way of implementing vaccination using local contact information called acquaintance vaccination (AV). According to this strategy, a randomly picked node is asked to name a neighbour node to be vaccinated. Therefore, no knowledge of the node degrees and any other global information of network are required. In fact, it selects the node that has a large number of connection to other nodes. Its efficiency greatly exceeds that of random vaccination. The acquaintance method is also improved in few other ways. Instead of vaccinating random acquaintance, it is more effective to vaccinate the acquaintance who has more frequent contact~\cite{deijfen2011epidemics}. That means the selected individual should be asked to name friends who contact frequently. This method substantially improves the efficiency of AV strategy. The works of ~\cite{holme2004efficient, chen2015improved,gallos2007improving} have shown that if neighbour nodes with many connections are vaccinated then the performances are improved significantly.

The acquaintance based strategies still require a large number of individuals to be vaccinated to achieve control goal. Thus, there have been a fair amount of works to search for other contact properties that can be obtained locally. The analysis of social contact networks shows that individuals are connected to various communities~\cite{girvan2002community,guimera2003self}. These properties of the contact network are exploited by several works where the concept of the bridge nodes is introduced as these nodes provide the pathways for a disease to propagate from one community to another community. Therefore, vaccinating such nodes will be a more effective strategy than of selecting random acquaintance. However, it requires the searching methods that use only the local contact information. The searching algorithm developed by the work of ~\cite{gong2013efficient} can find the bridge nodes using stochastic searching methods that need only local structural information. They found that the developed strategy based on these bridge nodes is more efficient than random strategies. A similar approach is applied in ~\cite{salathe2010dynamics} where bridge hub nodes, nodes bridging between two communities, are chosen for vaccination.

The above vaccination strategies have been developed based on static network properties. However, the real-world social contact networks are dynamic which has a strong impact on disease spreading and hence designing a vaccination strategy. The study~\cite{mastrandrea2015contact} shows that the contact rates between a pair of nodes are broadly distributed. Therefore, the selection of an acquaintance in AV strategy is not sufficient to find the appropriate nodes to be vaccinated. For example, the infection risk for being in a contact with an infected individual is relevant to the contact duration. Moreover, there is a higher risk if one susceptible interacts frequently with the infected individual. The works of ~\cite{lee2012exploiting,starnini2013immunization,masuda2013predicting} consider this information in neighbour selecting instead of selecting random neighbours. The authors of~\cite{lee2012exploiting} use the most recent contact for vaccination and they also apply weight to capture the contact rates with the neighbouring nodes. Theses vaccination strategies outperform acquaintance vaccination.

The collection of contact information is often difficult. Thus, vaccination with detailed local contact information may be infeasible in real-world scenarios and lose the benefit of using the local contact information. There have been several other vaccination models based on individual movement behaviours where individuals contact properties are not consider explicitly~\cite{mao2009efficient}. Beyond the contact properties, the work of ~\cite{mao2010dynamic} consider the individuals who travel long distance for vaccination. The similar approach is taken by ~\cite{miller2007effective} where individuals who visit many locations are vaccinated. 

The above vaccination strategies shows that node ranking is conducted based on the contact information about neighbouring nodes. However, in the diffusion processes with indirect interactions, it is difficult to identify the neighbours contacted through indirect interactions. Thus, there is a need to understand the efficiency of strategies with indirect interaction and find the best strategy. The authors of ~\cite{shahzamal2020vaccination} addressed this issues to build vaccination strategies with the indirect links.

\section{Conclusion}
For modelling disease spread within a society, a proper infection risk assessment model and a proper contact network are required. This paper analysed a wide range of networks models and infection risk models. There are simple infection risk assessment models that capture practical situation through simple model parameters. The complex model can capture the temporal dynamics of individual interactions and estimate more accurate infection risk. But, it require granular level contact information which may not available all time. In addition, it is observed that many factors affect the spreading dynamics of contagious items on contact networks. However, it is clear that interaction pattern of individuals is one of the key factors in driving diffusion processes on contact networks. There have, therefore, been a wide range of efforts to understand and integrate the impacts of interaction patterns with diffusion modelling~\cite{mossong2008social,mikolajczyk2008social,hens2009mining,iribarren2009impact}. There are, however, still some critical factors to be addressed in constructing proper diffusion models that capture realistic contact patterns. In addition, current opportunities for gathering individual-level contact data have attracted the researchers to deep dive further in this field by looking at contact patterns at the granular level~\cite{starnini2014time,stehle2011simulation,huang2016insights,tatem2014integrating,shahzamal2018impact}. The models presented in this paper would provide a guideline to model infectious disease like COVID-19.

\section{Introduction}

\bibliographystyle{unsrt}
% Bibliography, in BibTeX format (the references.bib file)
\bibliography{references}
\end{document}